# Giant Thermomechanical Bandgap Engineering in Quasi-two-dimensional Tellurium

*Naveed Hussain[1,2,†,\*], Shehzad Ahmed[3†], Hüseyin U. Tep[4], Kaleem Ullah[5], Khurram Shahzad [6], Hui Wu[2], Maxim R. Shcherbakov[1]\**

[1] Department of Electrical Engineering and Computer Science, University of California Irvine, Irvine, CA 92697, USA

[2] State Key Laboratory of New Ceramics and Fine Processing, School of Materials Science and Engineering, Tsinghua University, Beijing, 100084, China.

[3] College of Physics and Optoelectronic Engineering, Shenzhen University, Guangdong 518060, P. R China

[4] Micro and Nano-Technology Program, School of Natural and Applied Sciences, Middle East Technical University, Ankara, 06800, Turkey

[5] Department of Electrical and Computer Engineering, University of Delaware, Newark, DE 19711, USA

[6] Department of Physics, Middle East Technical University, 06800 Ankara, Turkey

† These authors contributed equally

**Abstract**

Mechanical straining-induced bandgap modulation in two-dimensional (2D) materials has been confined to volatile and narrow modulation due to substrate slippage and poor strain transfer. We report the thermomechanical modulation of the inherent bandgap in quasi-2D tellurium nanoflakes (TeNFs) via non-volatile strain induction during hot-press synthesis. We leveraged the coefficient of thermal expansion (CTE) mismatch between TeNFs and growth substrates by maintaining a high-pressure enforced non-slip condition during thermal relaxation (623 to 300K) to achieve the optimal biaxial compressive strain of -4.6 percent in TeNFs/sapphire. This resulted in an enormous bandgap modulation of 2.3 eV, at a rate of up to ~600 meV/%, which is two-fold larger than reported modulation rate. Strained TeNFs display robust band-to-band radiative excitonic blue photoemission with an intrinsic quantum efficiency (IQE) of c.a. 79.9%, making it promising for energy efficient blue LEDs and nanolasers. Computational studies reveal that biaxial compressive strain inhibits exciton-exciton annihilation by evading van-Hove singularities, hence promoting radiative-recombination. Bandgap modulation by such nonvolatile straining is scalable to other 2D semiconductors for on-demand nano(opto)-electronics.

**Keywords**: *Hot-pressing, thermomechanics, bandgap engineering, strain induction, tellurium nanoflakes, 2D materials, blue photoemission*

**Introduction**

Thegapless nature of graphene,[1] has sparked a huge interest in discovering novel two-dimensional (2D) semiconductors with tailored bandgaps.[2] For the latter, introducing a slight disorder in lattice symmetry by mechanical straining may drastically influence the electronic configuration in 2D semiconductors, resulting in massive bandgap modulation. Thus, mechanical strain induction to modulate the bandgap of various 2D semiconductors remains highly anticipated for custom-designed optoelectronic and photonic devices operating at broad spectral ranges.[3] Typical epitaxial growth,[4] mechanical bending,[5-9] chemical doping,[10,11] 2D vdW stacking,[12,13] alloying[14,15] and applying high-pressure (~< 100 GPa),[16,17], integration with nanostructures substrates,[18] and wrinkling effect[19] have been notable approaches for inducing lattice strain. Most of these approaches are incompatible with the manufacturing of optoelectronic devices because they generate volatile strain in 2D materials and demonstrate only the transient bandgap tunability. On the contrary, the techniques that can introduce non-volatile/permanent lattice strain suffer from limited overall strain induction (<0.5%). The maximum modulation rate S ($\Delta E_g/\varepsilon$), which defines the efficacy of a straining technique, has been reported at around 300 meV/%.[20] Thus, the bandgap modulation has been restricted to a modest value of a few hundred meV's, which is insufficient to cover large spectral bandwidths.[21] Although a non-volatile compressive strain of -2.4% in millimeter-thick epitaxial perovskite films has been reported,[22] achieving massive non-volatile strain induction in ultrathin 2D nanomaterials to enable large bandgap modulation remains a challenge that needs to be addressed.

Group VI tellurium (Te) is a versatile van der Waals (vdW) material that has attracted significant scientific interest as a prospective new member of 2D family due to its ability to stabilize in a monolayer.[23] While the field effect transistors (FETs)[24] and black-body sensitive IR photodetectors[25] on solution-grown 2D Te has been reported, its narrow direct band gap of ~0.35 eV has severely restricted its applications in the visible and near-infrared optoelectronics.[26] A recent theoretical work predicted that Te is likely to undergo dramatic bandgap transition (0.35 to 1.92 eV)[27] via confinement effect[28] and mechanical straining[29], making it a promising candidate for nano-optoelectronics.[30] It has also been hypothesized that a compressive biaxial stain in 2D-Te would allow for a substantial bandgap modulation and robust absorption in ultraviolet-blue visible light, [31] which would ideally require permanent yet enormous bandgap modulation. So far, mechanical straining alone have shown to be ineffective, coupling it with heat treatment may constitute an ideal platform to achieve such bandgap modulation.

A mismatch between the linear coefficient of thermal expansion (CTE) of 2D materials as produced and the growing substrate can generate stresses of up to 0.2%.[32,33] In contrast, the CTE mismatch may be utilized more effectively when coupled with a large uniaxial (out-of-plane) pressure that maintains a non-slip condition between 2D materials and substrate during their thermal expansion and relaxation.

We synthesized quasi-2D Te nanoflakes (TeNFs) directly from bulk tellurium and concurrently obtained massive bandgap modulation from 0.35 to 2.65±0.1 eV, caused by non-volatile biaxial compressive strain using a hot-pressing technique.[34,35] Micro-Raman and high-resolution transmission electron microscope (HRTEM) investigations validated the induction of substrate-induced non-volatile strain in TeNFs synthesized on c-sapphire ($Al_2O_3$), indium tin oxide (ITO) glass, and silicon (Si) substrates. TeNFs synthesized on sapphire (TeNFs/sapph.) demonstrated the largest strain-tuned bandgap modulation of 2.3 eV and bright blue photoemission, as characterized by UV-visible absorption and continuous-wave (CW) photoluminescence (PL) emission spectroscopy. Lifetime PL measurements performed on TeNFs reveal a long radiative lifetime of 2.2 ns with a high intrinsic quantum efficiency (IQE) of 79.9 %. This work provides an innovative insight in the field of strain-induced band gap modulation and contributes to broadening the scope of blue LEDs and photodetectors.

**Synthesis and strain estimation using lattice vibrational characterizations of TeNFs**

Figure 1a shows a typical hot-pressing technique in which agglomerates of tellurium microparticles (MPs) are molded to quasi-two-dimensional (2D) flakes by mechanical squashing at a high temperature (350 °C). Te-agglomerates were subjected to about 1 GPa of out-of-plane uniaxial pressure, restricting their thermal expansion solely to the x-y plane.

A thorough synthesis mechanism is described in the text S1, supporting Information. Fig. 1b presents a field emission electron microscopy (FESEM) image of a typical quasi-2D trapezoidal Te flake. 2D Te preferentially adopts a trapezoid shape due to its 1D preferred growth along its chiral-chain lattice, which spirals along the [0001] direction, making it easier to distinguish between the *a (zigzag)* and *c (armchair)* axes.[36] The atomic force microscopy (AFM) image showing a 12-nm thick TeNF is presented in Fig. 1 (c). The optical microscope image of the same TeNF is displayed in the upper-left inset. We used both pressure and heat to achieve a non-volatile biaxial compressive strain in the TeNFs crystal lattice. During the transformation from MPs to NFs, the uniaxial pressure along the *z*-axis maintained the non-slip condition between the TeNFs and the bottom substrate. The key feature of our strain induction approach lies in the mismatch in the thermal expansion coefficient (CTE) between the TeNFs and the top and bottom substrates during both expansion and contraction processes to enable substrate-based biaxial lattice shrinkage, resulting in compressive strain induction. We employed a variety of substrates, including c-sapphire, ITO, silicon, to achieve control over the induced compressive strain. However, in the main text of this paper, we mostly focus on the results of TeNFs produced on sapphire, as they have shown to be the most dramatic. The optical microscopy, FESEM, and atomic force microscope (AFM) images of a TeNF fabricated on ITO substrate are shown in Figure S1 (a-c), supporting information. The energy-dispersive X-ray spectroscopy (EDS) spectra acquired from an ultrathin TeNF/ITO is shown in Fig. S1 (d-e), supporting information revealing high purity of as-prepared TeNFs. A crack propagating on ITO signifies the importance of high pressure in the fabrication of ultrathin TeNFs (Fig. S1 (e), supporting information). The thickness distribution histogram of 100 thinnest TeNFs on c-sapphire exhibited a significant percentage (60-65%) of flakes in the 10 to 30 nm thickness range, according to the AFM data (Fig. S4 (a), supporting information).

To quantify the biaxial compressive strain and resulting structural evolution in crystal lattice vibrations in TeNFs, we performed detailed micro-Raman (µ-Raman) investigations. Due to the anisotropic structure of Te, the polarized µ-Raman spectroscopy was performed to individually probe the structural fingerprints in both the *a (zigzag) and c (armchair)* axes. Because of the trapezoid shape of TeNF, the axes were easily identified in the µ-Raman studies. The Raman spectrum of the bulk Te typically consists of three active phonon modes $E^1$, $A^1$ and $E^2$, that are located at 89.9, 116.8 and 137.4 cm$^{-1}$, respectively.[37,38] The principal $A^1$ mode, represented by the most prominent Raman peak, is assigned to lattice vibrations parallel to the basal planes, whereas $E^1$ and $E^2$ are degenerate phonon modes primarily assigned to bond-bending caused by the rotation along *zigzag* and asymmetric band-stretching alongside the *armchair*, respectively.[39] A shift in $E^1$ and $E^2$ modes thus corresponds to the strain along the *zigzag* and *armchair* axes, respectively. As a result, we employed a linearly polarized 532 nm laser to quantify the directional strain in TeNFs. The polarization along the *zigzag* direction probes the lattice strain along the 1D chain direction, i.e. *armchair*; see the schematic in Fig. 1d.[40] Fig. 1e presents the comparative µ-Raman spectra of an unstrained bulk Te (black curve) and strained TeNFs lying on Si (blue curve), ITO (green curve), and sapphire (red curve) substrates. A distinct blue-shift confirmed the biaxial compressive strain along the zigzag direction in both $A^1$ and $E^2$ modes. The estimated blueshift in the $A^1$ phonon mode is plotted against the estimated compressive strain, and the estimated compressive strain in both the degenerate ($E^1$ & $E^2$) modes for sapphire, ITO, Si, and bulk Te is shown in Fig. 1 (f, g). The amount of strain in TeNFs and its type is estimated by using $\varepsilon(\%) = \frac{\omega_{A1}(bulk) - \omega_{A1}(2D)}{\omega_{A1}(bulk)} \times 100$, where $\omega_{A1}(bulk)$ and $\omega_{A1}(2D)$ are the peak values of $A_1$ phonon modes corresponding to bulk Te and TeNFs, respectively. The significant blue-shift in $E^2$ mode confirmed the compressive strain along the armchair direction in all the samples, while the negligibly small blueshift in $E^1$ mode (blue triangles in Fig. 1g), because the laser was polarized along the *zigzag* direction. We rotated the sample clockwise by 90° so that the laser was polarized along the *armchair* direction, allowing us to measure the strain along the zigzag (Fig. 1h). The μ-Raman spectra of TeNFs manufactured on c-sapphire, ITO, Si, and bulk sample were compared. They revealed a prominent blue shift in $E^1$ phone mode, as well as a blue-shifted $A^1$ mode, with negligible blue-shift in $E^2$ (Fig. 1i). Fig. 1j shows the amount of compressive strain in response to the blue shift in $A^1$ phonon mode, whereas Fig. 1k depicts the blue shift and the corresponding c-strain in $E^1$ and $E^2$ modes. In comparison to the significant strain along the *zigzag* direction, there was minimal compressive strain along the *armchair* because of the inability of the setup to probe the strain along the armchair direction. Using the $A^1$ phonon mode, the optimal value of compressive strain estimated along the *armchair* for TeNFs/sapphire was -4.6±0.35%, while it was -4.250±0.25% along the *zigzag*. Meanwhile, for TeNFs/ITO samples, the strain along the armchair and *zigzag* was estimated to be -3.750±0.28% and -3.180±0.3%, respectively; see Table 1. We deconvoluted the μ-Raman spectra of both bulk Te and TeNFs/sapphire and performed peak fitting using Gaussian function to estimate the FWHM (Γ) of each vibrational mode (Fig. S2 (a,b), supp. info.). As compared to the bulk Te,

TeNFs/sapphire exhibited significantly reduced Γ values, which indicated that their improved crystallinity and substantial reduction in the native defect density.

To study the homogeneity of compressive strain in TeNFs/sapph, we performed Raman mapping of the phonon modes as shown in Fig. 1l. The bright spots represent the phonon modes at different points labeled as P1-P5. The inset shows the blue-shift Δω=-5.47 cm$^{-1}$ in the A$^1$ phonon mode. The µ-Raman spectra recorded from points P1-P5 (shown in the inset) are displayed in Fig. 1m, indicating the strong spatial localization of compressive stain in TeNFs. The inset shows the points distribution on the TeNF, where the scale bar is 5 µm. After establishing the strain localization, it was important to gauge the strain gradient along the Te flakes. The thermomechanically induced compressive strain gradient in TeNFs was validated using COMSOL Multiphysics simulations (Fig. 1n). The strain gradient at the TeNFs/sapph interface was predominantly introduced during the relative thermal contraction and is displayed from left to right in a systematic manner. The top row represents the strain gradient from above, while the bottom row depicts its cross-sectional view at the interface of TeNFs and sapphire. Due to their maximum thermal expansion state, the interfacial compressive strain is at its minimum when the hot press reaches its maximum temperature (350 ˚C), i.e., T$_{max}$.-T$_{room}$ is 10 ˚C. The CTE mismatch causes c-sapphire and TeNF to contract at dissimilar rates throughout the temperature relaxation to T$_{room}$, resulting in a strain gradient from the flake edges to the core at room temperature (T$_{max}$.-T$_{room}$ = 20 ˚C).[41] We calculated the relaxation constant: $R = (\omega_{max.strained} - \omega_{2D\ Te})/(\omega_{max.strained} - \omega_{bulk\ Te})$, where ω$_{max.strained}$, ω$_{bulk\ Te}$, and ω$_{2D\ Te}$ are the peak values (in cm$^{-1}$) of A$^1$ phonon modes of maximum strained TeNF, bulk Te (relaxed) crystal and that of the TeNF with specific thickness, respectively.[42] Our experimental results (described in text S4, and shown in Fig. S3 (a, b, &c), supporting Information) reveal that R is zero at a thickness of 12 nm, but it reaches unity at a thickness of 60 nm when dislocations occur, plastically relaxing the strain in their proximity.

**Strain estimation with electron microscopy**

We employed high-resolution transmission electron microscopy (HRTEM) as a direct approach to visualize compressive strain induction in TeNFs. Tellurium crystallizes to form a hexagonal lattice with every fourth atom exactly above another atom in its chain, resulting in an equilateral triangle when projected on a plane perpendicular to the chain direction. An infinite helical rotation at 120° along the longitudinal [0001] *armchair* (c-axis) covalently connects these atoms.[43] The weak van der Waals force along the *zigzag* direction binds all turns to hexagonal bundles in the radial direction. Fig. 2 (a, b) show, respectively, the transmission electron microscopy (TEM) and HRTEM images of a thicker flake reminiscent of unstrained bulk Te and provides reference values of its lattice constants along both the *zigzag* and *armchair*-axes. The HRTEM in Fig. 1b indicates a perfect 120° between the two axes, with lattice constants of 4.7Å and 3.8Å along *armchair* and *zigzag*, respectively.[19] Fig. 2 (c-e) and Fig. 2 (f-h) consist of low- and high-magnification HRTEM images, and the corresponding selected area electron diffraction (SAED) pattern of TeNFs/sapphire and TeNFs/ITO samples, respectively. Fig. 2c presents the low-resolution HRTEM image, while the inset shows the TEM image of the TeNF. Fig. 2d shows a magnified HRTEM image collected from the area highlighted by the red box in Fig. 2c. A distinctive compression along both axes was observed, which we measured using the Gatan digital micrograph (DM3) software. The lattice constants shrank by about 0.16±0.35 Å (3.8 Å to 3.64±0.35Å) along the *zigzag* and 0.19±0.6Å (4.7Å to 4.51±0.6Å) along the *armchair* axes. Fig. 2f shows the TEM image (inset) and low-resolution HRTEM image of TeNF/ITO. The HRTEM image (Fig. 2g) shows the shrinkage of about 0.13±0.45Å (3.8Å to 3.67±0.45Å) along the zigzag and 0.17±0.5Å (4.7Å to 4.53±0.5Å) along the armchair axes. Furthermore, the nanaoscale biaxial compression also reduced the angle between the two axes from 120° to around 110° for TeNFs/sapphire (Fig. 2d) and 112° for TeNFs/ITO (Fig. 2g). The SAED patterns collected from both the samples also confirmed the angular compression between the axes, meanwhile revealing highly crystalline nature of TeNFs fabricated on both the substrates. The diffraction points along *zigzag* are visible, making it easier to determine the other axis.

The lattice parameters of bulk tellurium along the *zigzag* and *armchair* axes can be determined at various temperatures between 25° and 350° C by using the following equations: $a_{Te} = a_o + 168.3 \times 10^{-6} t - 12.68 \times 10^{-8} t^2$ and $c_{Te} = c_o - 14.99 \times 10^{-6} t$, respectively.[44] According to these equations, heating Te to 350 °C alone without maintaining the non-slip condition between the substrate and Te i.e. the absence of uniaxial pressure would only induce a strain of ~0.166%, reducing the lattice constant along the armchair ($a_c$) from 3.8 to 3.793 Å. Meanwhile, the lattice constant along the zigzag ($c_a$) would only reduce from 4.7 to 4.688 Å, at a strain rate of 0.25%. This shows that the combination of uniaxial pressure along with heating was inevitable for achieving such high strain values in TeNFs. We compare the compressive strain values obtained by two independent methods in Fig. 2i. The green and light-yellow lines with error bars represent the compressive strain in TeNFs obtained by μ-Raman and HRTEM techniques, respectively. Within the experimental uncertainties, the results are in striking agreement . The results also suggest that the overall strain in TeNF/ITO was quantitatively less than that of TeNF/sapphire. The inset shows evidence of nanostraining in TeNF/sapphire in the form of wrinkles. The ESD mapping spectrum showing a strong Te signal is presented in Fig S4b, supporting information.

**Optical determination of bandgap modulation and exciton dynamics in TeNFs**

Strain induction has a significant impact on the band structure of TeNFs. To investigate potential changes in band structure from the biaxial compressive strain, we performed both the CW and time-resolved photoluminescence (PL), absorbance, and emission spectroscopy characterizations over a single 12 nm thick TeNF using a variety of excitation wavelengths spanning from UV to NIR. Fig. 3a illustrates the absorbance (green dashed line) and µ-PL emission spectra (blue solid line) of TeNF and bulk Te (black line) excited by a 325 nm CW laser. Due to the narrow bandgap of 0.35 eV, the bulk Te showed virtually no PL response in the UV-IR region, apart from a weak line at around 1.9 eV attributed to the native defect states. TeNFs, in contrast, exhibited a strong PL emission signal in the blue region of the spectrum, peaking at 2.62 eV, confirming an enormous widening of bandgap ($E_g$) to blue spectral region. The optical $E_g$ of 2.65 eV was determined by extrapolating the tangent line of the absorbance measurements using the Tauc-plot, see the inset in Fig. 3b. This is consistent with the PL emission peak energy of 2.62 eV and suggests a large $E_g$ shift of 2.3 eV caused by the compressive strain. Strong blue photoemission in TeNF/sapphire in Fig. 3c provides compelling evidence of $E_g$ modulation, as measured by a fluorescence microscopy imaging technique. The deconvolution of the TeNF/sapph. PL map of the same TeNF is given in Fig. S5 (a, b), supporting information, showing spatial PL localization in TeNFs. The PL spectrum of TeNFs/sapphire consisted of a triplet state in the visible part of the spectrum (Fig. S 6a, supporting Information). The red-highlighted peak $X^0$ at 2.62 eV corresponds to the band-edge recombination of electron-hole (*e-h*) pairs.[45] The origin of the $X^D$ peak (highlighted in green), centered at 1.78 eV could be attributed to radiative *e-h* recombination due to the native defect states, as indicated by the peak's large width. However, the origin of the peak $X^{Int.}$ at 1.3 eV remains unknown. We hypothesize that this may be attributable to either the loosely bound surface oxide species on TeNF surfaces or the emission resulting from the interfacial states of tellurium and Sapphire. To provide a reasonable explanation for the genesis of this PL peak, however, further investigation is required. We excited TeNFs/sapph. with a variety of excitation wavelengths, including 325nm, 410 nm, 532 nm, 633 nm, and 785 nm, to investigate the interband transitions. We obtained a PL emission peak in response to all the excitation sources, which indicated the opening of sub-band states within the modulated bandgap of TeNFs (Fig. 3d). However, the intensity of the PL reduced significantly at longer excitation wavelengths, as shown in given in Fig. S6b, Supporting Information. The schematic illustration of interband transitions in TeNFs have been presented in Fig. 3e. The substrate-dependent comparative PL emission spectra of TeNFs/sapphire (blue), TeNFs/ITO (orange), TeNFs/Si (wine), and bulk Te (cyan) is shown in Fig. 3f. It is evident that ITO substrate also played a key role in significant bandgap opening in TeNFs, however, it had other enhanced and broad peaks that include a contribution from the native defects and, potentially, the TeNFs/ITO interfacial states. We performed thickness-dependent PL studies which revealed a monotonic decrease in PL intensity (~two orders of magnitude) with decreasing flake thicknesses from 200 to 10 nm (Fig. 3g). The reason for the drastic reduction in PL is the reduced absorption and, as a result, the reduced number of *e–h* pairs created by the excitation laser. This dependence is well described by the following expression: $I_{PL} = I_{Te}[1 - \exp(-a_{Te}h_{flake})]$, where $I_{Te}$ is the PL intensity of the TeNF with $h_{flake} = 200 nm$.[46] The absorption coefficient $a_{Te}$= 5000 cm$^{-1}$ was used.[47] We also observed an overall blue-shift of 170 meV in the PL peak energy with decreasing flake thicknesses from 200 to 10 nm, which is a signature of quantum-size effects in strained TeNFs. Fig. 3h reveals the variation in PL peak energy with flake thicknesses, while the variation in PL peak intensity with flake thicknesses has been presented in Fig. S6 (c, d), Supporting Information. The PL peak energy trend with the CTE of growth substrates has been presented in Fig. S 6e, Supporting Information. No contribution to the PL signal from the sapphire substrate, which usually has a peak at 1.79 eV, is observed (Fig. S7, Supporting Information). To exclude the possibility of oxidation, we performed comprehensive X-ray photoelectron spectroscopy (XPS) studies (Text S3 and Fig. S8, Supporting Information), manifesting a pure metallic phase with no oxide signatures.

Figure 3i shows the time-resolved PL traces of bulk Te and TeNFs, collected under ambient conditions using the time-correlated single photon counting (TCSPC), with an excitation wavelength of 325 nm. The PL lifetimes were calculated using a bi-exponential fitting equation and are important for understanding the exciton dynamics in TeNFs.[48] In comparison with the PL decay curve of bulk Te (inset), the decay curve of TeNFs showed enhanced carrier recombination lifetimes, where the shorter lifetime component ($\tau_1$) increased threefold, from 0.21 to 0.73 ns. Similarly, the longer lifetime component ($\tau_2$) almost doubled from 1.39 to 2.84 ns. This, together with the strong PL from TeNFs, suggests a reduced non-radiative recombination component in TeNF, caused by the bandgap opening. The carrier recombination lifetime for TeNFs was determined to be 2.22 ns by using the equation: $\frac{1}{\tau} = \frac{1}{\tau_{nr}} - \frac{1}{\tau_r}$, where τ, τ_r , τ_nr are the average, radiative and non-radiative lifetimes, respectively. The avg. carrier recombination lifetime of bulk Te, however, was calculated to be 1.2 ns, which is ~50% decrease. The internal quantum efficiency, IQE (η) of TeNFs was calculated by using the following equation:[49]

$$IQE_{PL} = \eta = \frac{1/\tau_r}{1/\tau_r + 1/\tau_{nr}} = \frac{1/\tau_r}{1/\tau} = \frac{\tau}{\tau_r} \qquad (1)$$

The η of TeNFs was calculated to be 79.9%, suggesting a strong radiative recombination process in photo-generated excitons. The compressive strain causes the splitting between the Van-Hove singularity (VHS) resonance and the exciton transition energy, resulting in considerable suppression of exciton-exciton annihilation (EEA) and promoting the radiative recombination process.[50]

**Calculations of strain-induced band gap modulation**

Density functional theory (DFT), as implemented by the Vienna ab initio simulation program (VASP), was used to perform electronic and optical calculations[51] for Te valence orbitals and electrons for pseudoatoms were $5s^2$ and $5p^4$, respectively. The compressive strain on monolayer Te is applied by changing the cell parameters (a and b, equally), as shown in Fig. 4a. The in-plane xy atomic orbitals overlap significantly more when additional compressive strain is applied, which has a major impact on the inner energy states energy shift. As illustrated in Fig. 4 (b), variation for the band structures of tensile-strained monolayer Te is the reduction of the band gap, while with the compressively strained monolayer shows the band gap increase. This agrees with previous calculations on the variations of the band gaps as functions of strained mono-layer and bi-layer $MoS_2$.[52,53]

A distinct pattern for the band gap shifts is seen when compressive strain is applied. When applied to the monolayer Te, the compressive strain has the opposite effects on the band structure, reducing the band gap while increasing under compressive strain. Additionally, under compressive strain, monolayer Te within the range of 0 to -4.6%, respectively. From the DOS, one can calculate a joint density of states $J(\omega)$ to reveal the origin of the difference in absorption between unstrained and strained phases (neglecting oscillator matrix elements and excitonic effects). The basis of the shift in the absorption of these phases is finally revealed by a detailed analysis of the components of the optical absorption spectra—the joint density of states (Fig. 4c). The electron localization function of both the unstrained and strained TeNFs, and the corresponding absorption spectrum have been presented in Fig. S9 (a, b), supporting Information.

**Bandgap modulation, modulation rate, and long-term strain retention in TeNFs**

Fig. 5a presents the bandgap modulation in response to the compressive strain introduced along both axes in TeNFs fabricated on various substrates. It shows that even though the strain achieved in TeNFs/Si is roughly around -1%, causing the bandgap modulation up to ~1.9% is not enough to cause the indirect to direct bandgap transition, as evident by the absence of the PL emission. The shaded gray area roughly corresponds to the region with no emission , despite significant bandgap modulation. However, strong PL emissions were recorded for TeNF/ITO and TeNFs/sapphire, indicating the crossover to the emission zone in modulated bandgap. It is also evident that the compressive strain in not homogenous along both the axes for both the samples, where the strain along *armchair* is higher than that of *zigzag* due to the weak vdW interaction.[19] Fig. 5b describes the bandgap modulation (ΔE) in response to the strain type and its values for various 2D material systems previously reported. The red dots represent the induction of volatile or short-lived strains, while the yellow stars present the induction of non-volatile or permanent strain engineering in 2D materials. The negative and positive strain values represent the induction of compressive and tensile strains, respectively. According to our knowledge, the maximum value of non-volatile strain reported in black phosphorus (BP) provides the overall ΔE of only 600 meV, while we have achieved ΔE of 2.31 eV for TeNF/sapph. and 2.25 eV for TeNF/ITO. Modulation rate S (ΔE), which is defined as the bandgap modulation per percent strain (meV/%), is the measure of the efficiency of the straining mechanism. Fig. 5c shows a comparative overview of the modulation rate for various reported 2D materials. The maximum S(ΔE) reported so far is ~300 meV/%, achieved in $MoS_2$ using the gated piezoelectric substrate, which is a non-sustainable strategy.[20] Whereas, by using the hot-pressing synthesis of TeNFs, we have achieved S(ΔE) of 600 and 502 mev/% on TeNF/ITO and TeNFs/sapphire, respectively. To investigate the long-live strain retention, we performed recurring Raman studies on TeNFs/sapphire samples up to 10 weeks after their fabrication with a step of two weeks. The blue shift in the $A^1$ vibrational mode remained the same after 10 weeks, which confirmed the induction of inherited non-volatile strain in as-prepared TeNF. We further performed the time-dependent PL studies presented in Fig. S10, supporting information. After one week of ambient environment exposure, there was a 31% loss in PL intensity as compared to the freshly prepared TeNFs/sapphire samples due to the photobleaching caused by surface degradation.[54] However, the PL intensity remained the same for weeks 2–10, indicating high optical quality and bandgap retention in TeNFs.

**Conclusion**

In this work, by means of a simple hot-pressing method to thermo-mechanically mold Te nanoparticles into ultrathin Te nanoflakes, by sandwiching them between c-sapphire, ITO, and Si substrates. Because of the significant adhesion assisted CTE mismatch between the substrate and TeNFs, the hot-pressed TeNFs endured compressive biaxial strain, resulting in persistent, substantial bandgap modulation (0.35-2.65 eV). Quantitatively, an optimum strain of -4.6% in TeNFs/sapphire was obtained, whereas the bandgap modulation rate in TeNF/ITO was as high as 600 meV/%. The strong blue photoemission in TeNFs is due to strained-induced suppression of exciton-exciton annihilation (EEA), which enhanced radiative carrier recombination. This report outlines hot-pressing as an efficient technique for introducing hereditary bandgap engineering in quasi-2D materials during their synthesis by manipulating the substrate-induces lattice strain.

**METHODS**

**Materials**

Commercially available Te microparticles (TeMPs) with size distribution of 10 to 50 µm (Sigma Aldrich, 99.99%) are used without any processing, to prepare an ethanol/ TeMPs (50mL/10mg) dispersion. Atomically flat (10×10 mm$^2$) c-cut sapphire ($Al_2O_3$) substrates (KYKY Technology Co. LTD) are ultrasonicated in acetone, ethanol, and distilled water. To obtain moisture-free surfaces of c-sapphire before mass-loading, the substrates are placed in a furnace maintained at 70 °C for 30 minutes. A commercial hot-press system (AH-4015, 200 V-20A, Japan) is used for thermo-mechanical squashing of TeMPs to fabricate ultrathin TeNFs.

**Fabrication of Polymorphic TeNFs**

Mass-loading was performed by drop-casting the 20 µL of the homogeneous ethanol/TeMPs dispersion on to the c-cut sapphire substrate by using a micropipette. The loaded substrate is left to dry for 30 minutes in an Argon filled glove box (MB200MOD) to avoid unnecessary exposure to moisture and dust particles. After the complete evaporation of ethanol, the c-sapphire substrate capped with large agglomerates of TeMPs of different sizes and shapes is taken out of the glovebox and is further covered by another sapphire substrate thus sandwiching the TeMPs. This sandwiched assembly (sapphire-TeMPs-sapphire) is then placed at the middle of the metal plates of the hot-press machine. During thermo-mechanical squashing, the temperature of hot-press is raised from RT to 350 °C, while the pressure is gradually increased from the atmospheric pressure to 1 GPa. The sapphire-TeMPs-sapphire assembly is hot-pressed for 30 min at 350 °C before allowing thermal relaxation to RT, while maintaining a constant pressure of 1 GPa throughout the relaxation process. The obtained sample is studied without any post-fabrication annealing.

**Material characterizations**

Morphology of TeNFs is studied by using the field emission scanning electron microscopy (FESEM), (MERLIN VP compact, Carl Zeiss, Germany). A multipurpose JEOL-2100 analytical electron microscope with the resolution of 0.19 nm, operated at 200 to 80 kV, was used to perform the transmission electron microscopy (TEM), high-resolution transmission electron microscopy (HRTEM) and selected area electron diffraction (SAED) measurements. To prepare TEM samples, c-sapphire capped with hot-pressed TeNFs were immersed in ethanol and ultra-sonicated for 30 minutes to isolate TeNFs from sapphire into the ethanol solution. The ethanol containing SeNFs is dropped onto the carbon-coated Cu grid by a micropipette to perform TEM/HRTEM characterizations. For scanning probe microscopy characterizations, an Asylum Research Cypher AFM (SPM, SHIMADZU Corporation, spm-9600) in tapping mode is employed to investigate surface topography and thicknesses of TeNFs on sapphire and ITO. A Raman spectrometer (LabRAM HR Evolution, HORIBA Jobin Yvon, France) with an objective lens (Nikon Plan Fluor 50 X, N.A.=0.4) and the laser spot size of approximately 2 μm is deployed to acquire Micro-Raman spectra of stained TeNFs lying an c-sapphire. X-ray photoelectron spectroscopy (XPS, Escalab, 250 Xi, Thermo Fisher Scientific, MA, USA), equipped with an A1Kα radiation source (1487.6/eV) is used to investigate the chemical composition and stoichiometry of TeNFs. The binding energy calibration is performed carefully by using C1s peak (284.8 eV) as reference value. PL measurements of TeNFs from RT to cryogenic temperatures with variable laser power ranging from 0.2-1 mW are performed by using LabRAM HR Evolution, HORIBA Jobin Yvon, France, equipped with a 532 nm Nd: YAG (Nd: $Y_3Al_5O_{12}$) laser, and a CCD detector.

**DFT Calculations**

first-principles calculations are performed using the projector augmented wave method as implemented in Vienna ab-initio simulation package (VASP)[55] coupled with generalized gradient approximation (GGA) electron-electron interaction defined as Perdew-Burke-Ernzerhof (PBE) exchange-correlation functional. To prevent interactions between layers brought on by the periodic boundary condition, a vacuum layer with a thickness of ~25 Å along the z-direction. A plane-wave basis set with a kinetic energy cutoff of 650 eV and Perdew-Burke-Ernzerhofer (PBE) projected augmented wave pseudopotential is adopted to expand the wave functions to describe the electron ion potential, respectively.[56] For structural relaxation, the Monkhorst-Pack method was used to mesh the reciprocal space at 21 × 19 × 1. To get a more precise band gap, the screened hybrid functional approach at the HSE06 level is also used.[57] The weak vdW interaction between the two monolayers is described by the DFT-D2 correction of Grimme.[58] All atoms were relaxed with predicted forces of less than 0.01 eV Å$^{-1}$ and total energies of less than $10^{-6}$ eV/atom. Optimized mono layer Te unit cell with lattice parameters a = 4.48 Å, b = 4.5.98 Å and c = 25 Å were considered.[59-61] The lattice parameters were slightly larger while in good agreement with the available experimental results.

**Pressure Calculation**

The pressure applied during hot-press for mechanically squashing TeMPs to fabricate ultrathin TeNFs is estimated by the following expression:

$$P = \frac{P_0 \times A_0}{a_0} = \frac{(55 \times 10^6 \mathrm{Nm^{-2}})(3.1416 \times (24 \times 10^{-3} \mathrm{m})^2)}{(1 \times 10^{-4} \mathrm{m^2})} = 0.996 \mathrm{GPa} \cong 1 \mathrm{GPa}$$

Where "$P_0$" is the applied pressure, "$A_0$" is the area of the cylinder, and "a" is the area of the substrate.

**Present Addresses**

[1]Department of Electrical Engineering and Computer Science, University of California Irvine, Irvine, CA 92697, USA

[2]State Key Laboratory of New Ceramics and Fine Processing, School of Materials Science and Engineering, Tsinghua University, Beijing, 100084, China.

**Corresponding Authors**

*Email: naveedh1@uci.edu

*Email: maxim.shcherbakov@uci.edu

**ACKNOWLEDGEMENTS**

**Funding Sources:** This study was jointly supported by the National Key Research and Development Program (No. 2019YFB2203400), the National Basic Research of China (Grant No. 2018YFB0104404) and National Natural Science Foundations of China (Grant No.12150410313).

**Author contributions:** N.H. initiated the project after consultation with H.W. N.H. synthesized and performed structural opticla characterizations of TeNFs. A.S. performed HSE calculations to analyze electronic structure modulation with bandgap and found theoretical evidence of strong photoemission. H.T. performed COMSOL Multiphysics simulations for compressive strain gradient in TeNFs. N.H., M.S., and K.S. performed the data analysis. N.H. and A.S. wrote the manuscript with contributions from all other authors. N.H. and M.S. supervised the project.

**Competing interests**: The authors declare no competing financial interest.

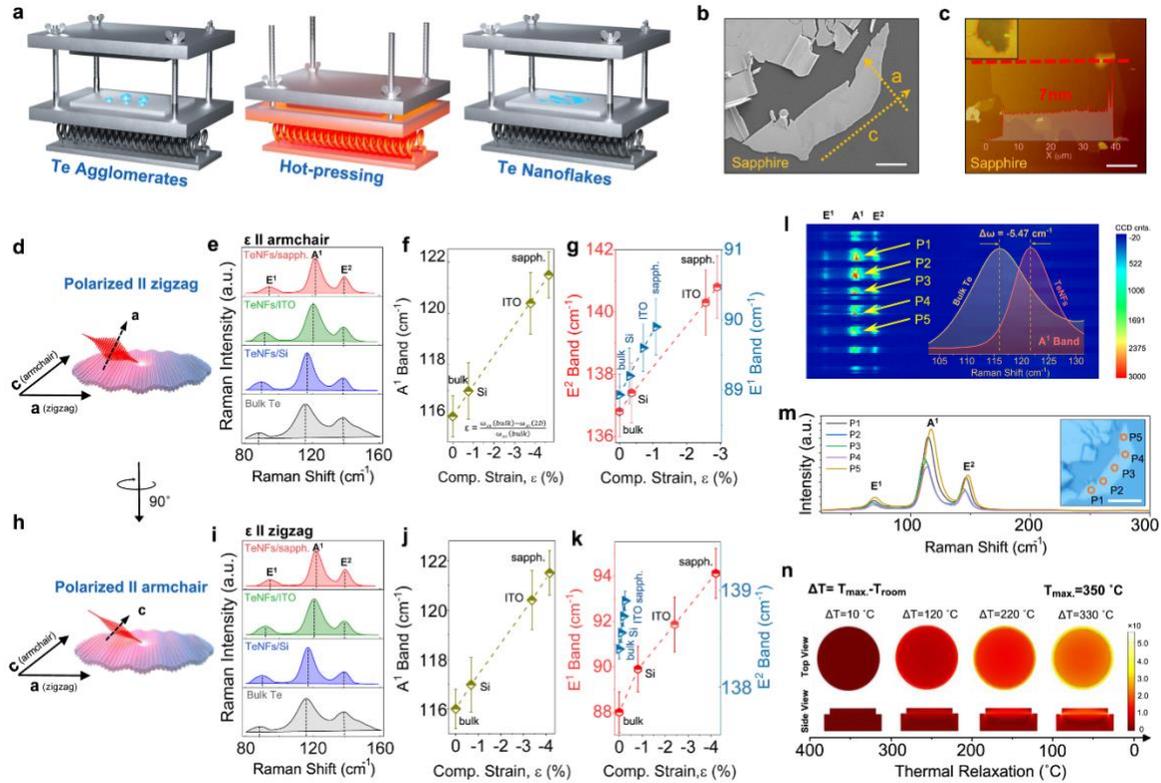

**Fig. 1**. **Hot-pressing strategy and μ-Raman spectroscopy characterizations of TeNFs.** (**a**) A schematic diagram of hot-pressing method that mechanically squashes TeMP agglomerates to mold them into ultrathin TeNFs. (**b**) FESEM image of a trapezoid-shaped TeNF lying on c-sapphire, with its zigzag (*a*) and armchair (*c*) axes. (**c**) Atomic force microscopy (AFM) image the TeNF/sapphire with the thickness profile of 7 nm. (**d**) Multilayer Te flake exposed to 532 nm laser polarized along zigzag (a-axis). (**e**) Comparative Raman spectra of TeNFs/sapph., TeNFs/ITO, TeNFs/Si, and bulk Te (**f&g**) Blue shift in $A^1$ mode with and combined $E^1$ and $E^2$ modes with compressive strain to estimate strain along armchair axis. (**h**) Te multilayer exposed to 532 nm laser polarized along armchair (c-axis). (**i**) Comparative Raman spectra of TeNFs/sapph., TeNFs/ITO, TeNFs/Si, and bulk Te. (**j&k**) Blue shift in $A^1$ mode with and combined $E^1$ and $E^2$ modes with compressive strain to estimate strain along zigzag axis. (**l**) Raman map of an individual TeNF and its the intensity profile confirming the presence of $E^1$, $A^1$, and $E^2$ phonon modes. (**m**) Raman spectra acquired from points P1-P5, showing strain localization. (**n**) COMSOL multiphysics simulation results showing the theoretical calculation of compressive strain

and its gradient at the sapphire-TeNF interface during the thermal relaxation process in hot-pressing.

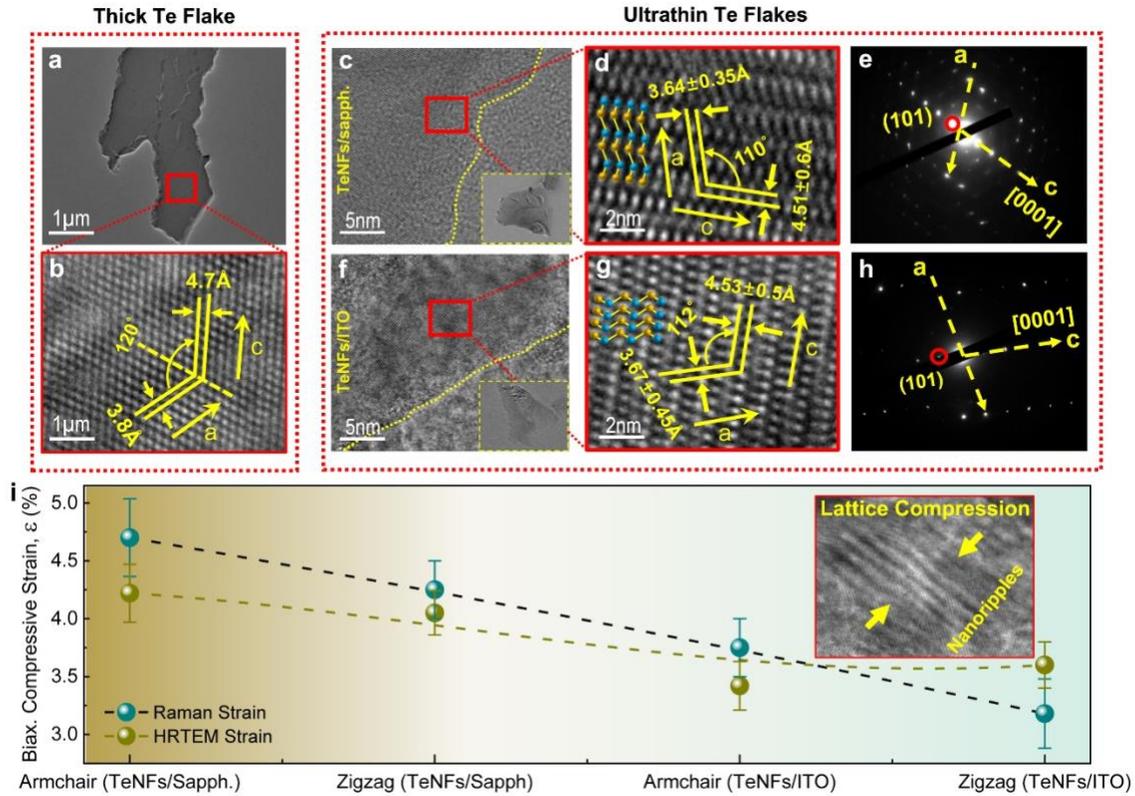

**Fig. 2**. **Comprehensive electron microscopy for strain estimation.** **(a)** Transmission electron microscopy (TEM) image of a thick TeNF reminiscent to bulk Te. **(b)** HRTEM image acquired from the highlighted by red rectangle in Fig. 1a shows standard lattice parameters along zigzag and armchair. **(c)** Low magnification HRTEM image acquired from the TEM image shown in the inset for TeNFs/sapph. sample. **(d)** High resolution HRTEM image showing lattice compression along both the axes. **(e)** Selected area electron diffraction (SAED) pattern, which exhibits highly crystalline nature of TeNFs/sapph. **(f)** Low magnification HRTEM image acquired from the TEM image shown in the inset for TeNFs/ITO sample. **(g)** High resolution HRTEM image showing lattice compression along both the axes. **(h)** Selected area electron diffraction (SAED) pattern, which exhibits highly crystalline nature of TeNFs/ITO. **(i)** The agreement between the strain values estimated using micro-Raman spectroscopy (green line), and by electron microscopy (TEM/HRTEM) (light yellow) along both the axes in TeNFs/sapph. and TeNFs/ITO

samples. The error bars were acquired by repeating each measurement 5 times. The inset shows the evidence of nanostraining in the form of nanoripples in TeNFs.

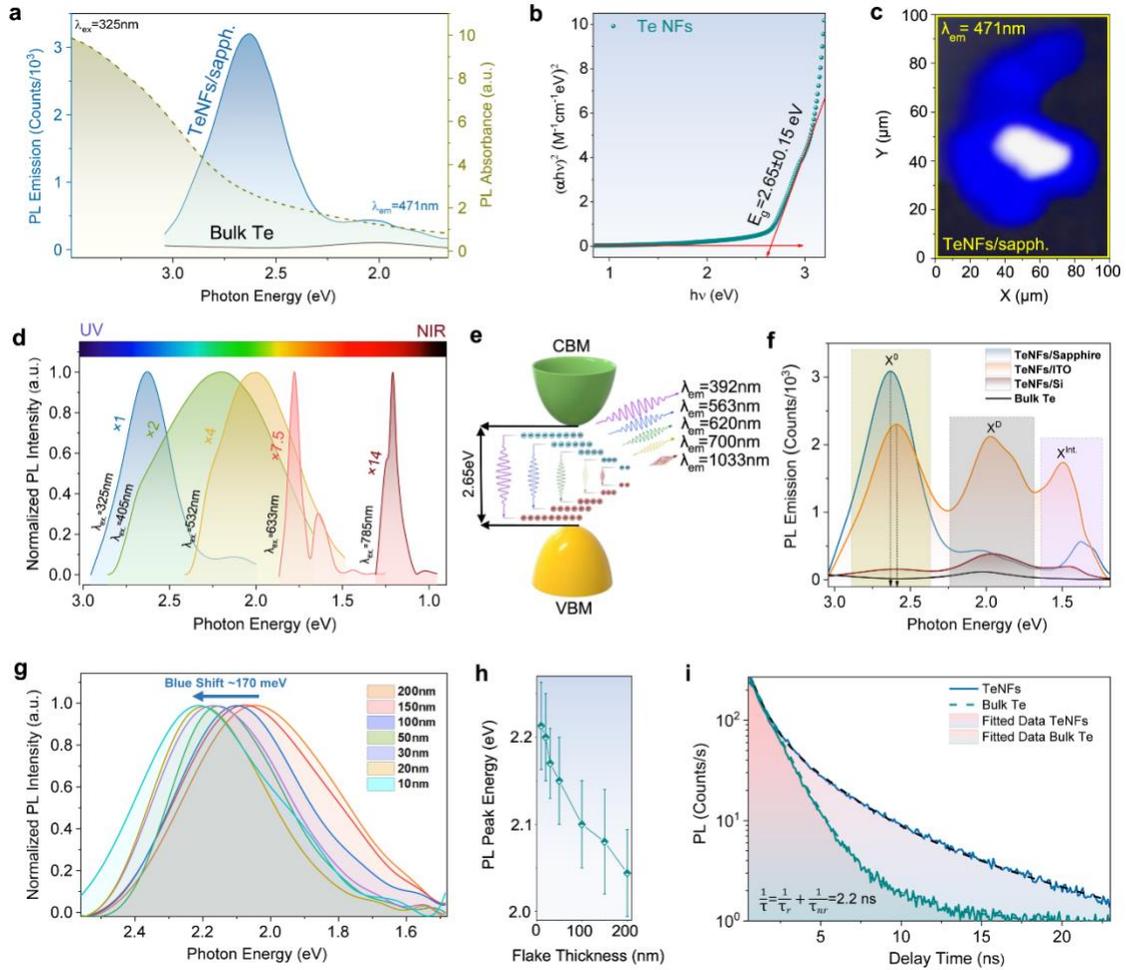

**Fig. 3**. **Optical determination of bandgap modulation and exciton dynamics in TeNFs.** **(a)** PL absorbance and emission spectra of TeNFs/sapph. acquired at 300 K, using 325nm laser. The inset shows the Tauc plot for band gap estimation, while another inset is the fluorescence microscopy image of an individual TeNF/sapph. exhibiting strong blue photoemission. **(b)** Excitation wavelength-dependent PL emission in TeNFs/sapph. **(c)** Schematic of sub-band transitions with different excitation wavelengths. **(d)** Substrate-dependent PL emission responses from TeNFs/sapph., TeNFs/ITO, TeNFs/Si, and bulk Te samples. **(e)** Thickness-dependent PL emission in TeNFs/sapph. The inset depicts the variation in PL peak energy with flake thicknesses. **(f)** Life-time PL emission spectra of TeNFs. The inset shows the comparative TRPL curve of bulk and TeNFs.

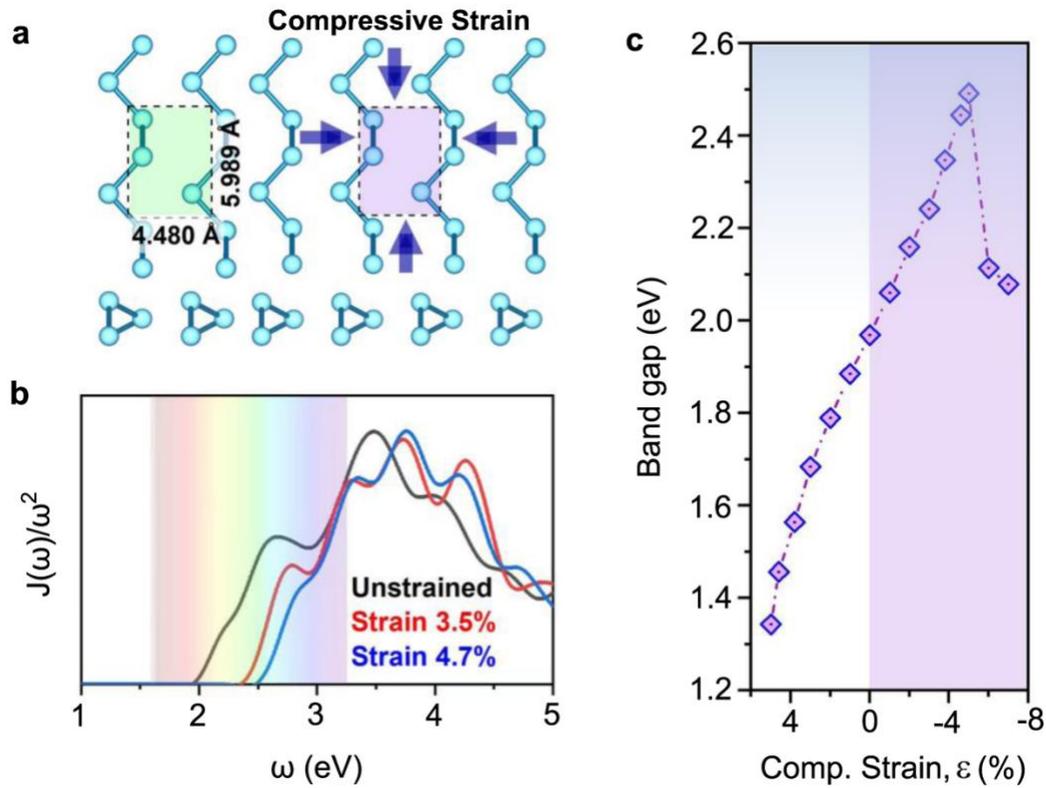

**Figure 4.** (**a**) Snapshots of layered Te, Top view and side view, the dashed lines show the primitive unit cell (with lattice vectors a-and b-). (**b**) Hybrid functional HSE06 predicted band gap effects on layered Te as a function of strain. Starting with the relaxed structure, the band gap experiences a transition with both tensile and compressive strain, which is in great agreement with that predicted by the standard DFT. Although DFT underestimates the band gap, its calculated gap-strain variation trend is consistent with that of HSE06. (**c**) HSE06 based Joint Density of States ($J(\omega)/\omega^2$) of the relaxed and strained Te layer respectively.

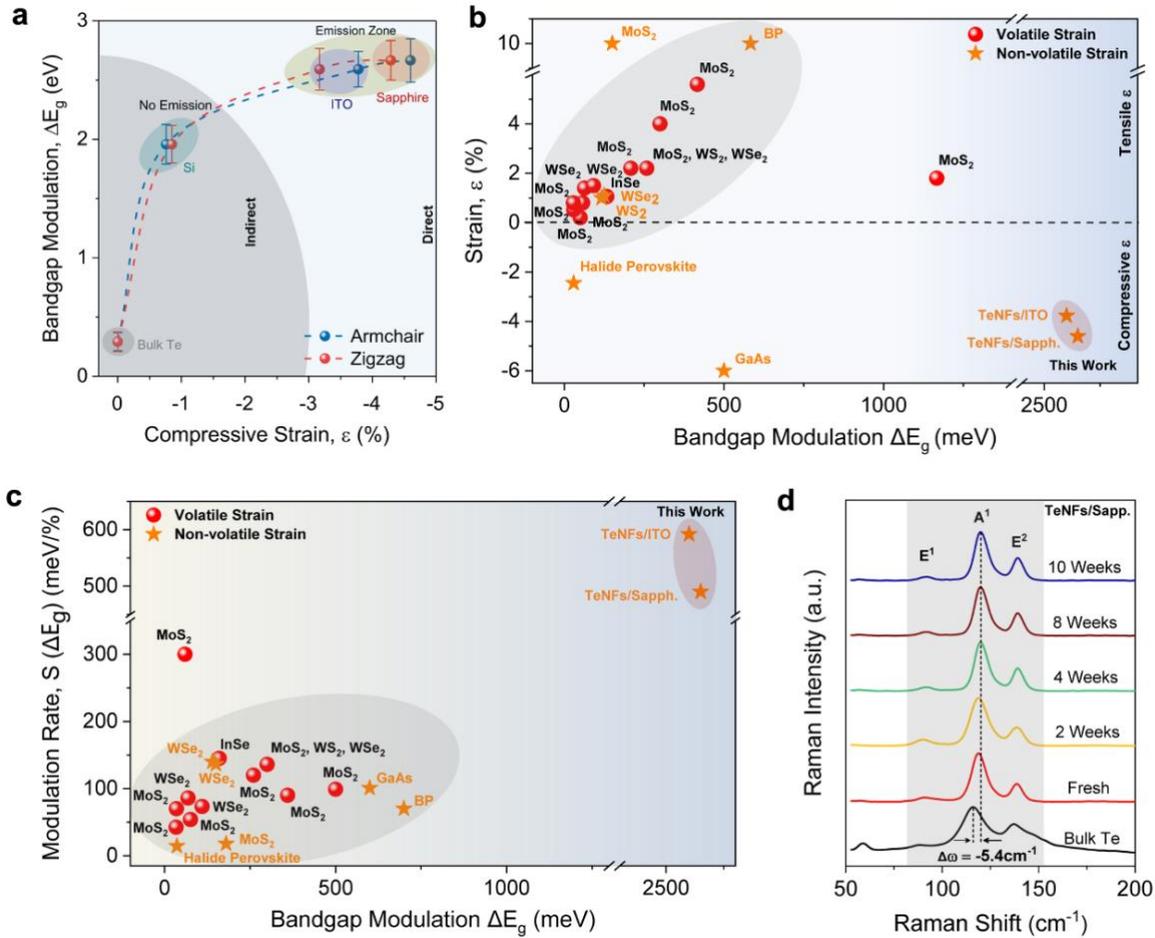

**Fig. 5. Bandgap modulation, modulation rate and long-term strain retention in TeNFs** **(a)** The overall bandgap modulation in TeNFs synthesized on Si, ITO, and sapphire substrates, in response to the compressive strain. **(b)** Previously reported overall bandgap modulation in 2D materials vs our work. The yellow star and red dots show non-volatile and volatile strain, respectively, while negative and positive strain values are associated with compressive ad tensile strain. **(c)** Modulation rate (ΔS) in response to bandgap modulation for previously reported works, where yellow stars represent works reporting non-volatile strain induction **(d)** Time-dependent µ-Raman spectra showing long term strain retention in TeNFs/sapph.

# Supporting Information

# Giant Thermomechanical Bandgap Engineering in Quasi-two-dimensional Tellurium

*Naveed Hussain[1,2,†,\*], Shehzad Ahmed[3†], Hüseyin U. Tep[4], Kaleem Ullah[5], Khurram Shahzad [6], Hui Wu[2], Maxim R. Shcherbakov[1]\**

[1] Department of Electrical Engineering and Computer Science, University of California Irvine, Irvine, CA 92697, USA

[2] State Key Laboratory of New Ceramics and Fine Processing, School of Materials Science and Engineering, Tsinghua University, Beijing, 100084, China.

[3] College of Physics and Optoelectronic Engineering, Shenzhen University, Guangdong 518060, P. R China

[4] Micro and Nano-Technology Program, School of Natural and Applied Sciences, Middle East Technical University, Ankara, 06800, Turkey

[5] Department of Electrical and Computer Engineering, University of Delaware, Newark, DE 19711, USA

[6] Department of Physics, Middle East Technical University, 06800 Ankara, Turkey

*Supporting Information contains the following content:*

**Text S1.** Detailed fabrication mechanism of strained TeNFs.

**Text S2.** Thickness-dependence of Relaxation Constant.

**Text S3.** Detailed XPS analyses of strained TeNFs

**Text S4.** COMSOL Multiphysics simulations for strain estimation.

**Text S5.** Detailed Computational Studies.

**Fig. S1. a,** An optical microscopy image of a TeNF fabricated on ITO substrate. **b,** Field emission scanning electron microscopy (FESEM) image of the same TeNF. **c,** Atomic force microscopy image of the identical TeNF with its height profile. The thickness was estimated to be 20nm along the blue dotted line. **d,** EDX map of TeNF acquired from Fig. S1e, showing the strong tellurium signal. **e,** An SEM image showing the crack propagation through a Te flake because of high pressure.

**Fig. S2. a,** A deconvoluted μ-Raman spectrum obtained from bulk Te. Peak fitting was performed to calculate the FWHM of each peak, which is numerically presented with in the corresponding peaks. **b,** A deconvoluted μ-Raman spectrum obtained from TeNFs/sapph.

**Fig. S3. a,** Thickness-dependent μ-Raman spectra showing consistent blueshift in $A^1$ and $E^2$ modes with decreasing thicknesses. **b,** Correlation between relaxation constant and compressive strain as a function of flake thicknesses. **c,** Direct relation between the compressive strain and relaxation constant, R.

**Fig. S4. a,** Thickness distribution histogram acquired by the AFM fata of 100 thinnest TeNFs. **b,** EDS map of TeNFs acquired using TEM.

**Fig. S5. a,** PL map of a TeNF shown in the Fig. 3a (main text). **b,** A 3D plot of the PL map showing the highest PL intensity from the center of the flake.

**Fig. S6. a**, Deconvoluted PL spectrum of TeNF/sapphire showing the peak consisting of three singlets ($X^0$, $X^D$, and $X^{Int}$.). The inset shows the schematic of inner-shall transitions during the radiative excitation recombination process. **b,** The variation in PL intensity with various excitation wavelengths. **c,** Thickness-dependent PL response of TeNFs with using

532 nm laser. **d,** A bar graph showing the quantitative variation in normalized PL intensity with flake thicknesses. **e,** The PL peak energies as a functions of substrate's CTE.

**Fig. S7.** The comparative PL spectra of bare sapphire and TeNFs/sapphire, confirming no background PL signal from sapphire at around 1.79 eV.

**Fig. S8. a,** Te3d scan of bulk Te and Te NFs fabricated on Si substrate. **b,** O1s spectra of as-prepared Te NFs. **c,** C1s spectra of as-prepared Te NFs. **d,** Survey spectra of bulk Te and as-prepared Te NFs. The inset shows the magnified spectra of Te NFs acquired from the area highlighted by red-box.

**Fig. S9. a**, Electron localization function (ELF) of unstrained and strained Te flakes. **b,** Absorption spectra of TeNFs, obtained by DFT calculations.

**Fig. S10.** Time-dependent of PL spectra of TeNFs/sapphire showing consistent PL response after 10 weeks of fabrications. This also confirms long-term bandgap retention in TeNFs.

*Text S1: Detailed Fabrication Mechanism of strained TeNFs*

We fabricated TeNFs by thermally assisted mechanical molding of agglomerates of tellurium microparticles (TeMPs) onto various substrates (c-sapphire, ITO, silicon), using hot-pressing method (schematic diagram is shown in Fig. 1a, main text). Tellurium chunks were thoroughly ground to obtain corresponding microparticles for making a Te/ethanol dispersion (1mg/25mL). 20µL of dispersion was drop-casted on highly polished and cleaned substrates with dimensions of 10mm×10mm×0.5mm, by the pipet gun in an Argon-filled glove box to maintain the lowest possible exposure to oxygen. The substrates, after drying, were left with non-uniformly scattered agglomerates of TeMPs, which were further capped with a dimensionally identical substrate to form a sandwich-like assembly, which was placed between the heating steel plates (jaws) of a commercial hot press. The substrates under the jaws of the hot press were heated from room temperature ($t_R$) to 350 °C, while the pressure was gradually elevated from atmospheric (uniaxial) pressure to a range of around 0.74 - 1 GPa, depending on the type and nature of the substrate. After 30 min of attaining the temperature of 350 °C, the whole assembly was thermally relaxed to room-temperature ($T_{room}$), and TeNFs with different lateral sizes and thicknesses were obtained on both the bottom c-sapphire, while a fraction of flakes was torn apart by the top substrate.

## Text S2: Thickness-dependence of Relaxation Constant (R)

We used µ-Raman studies to quantify the degree of plastic strain relaxation with flake thicknesses by finding the relaxation constant R, where $R$ ($0 \leq R \leq 1$) reflects the connection between the film thickness and the degree of plastic strain relaxation. When R equals one, the lattice structure is regarded completely relaxed, and when R equals zero, it is considered completely strained. Any value between 0 and 1 implies a moderately relaxed/strained structure.[1] We calculated R using the formula $R = (\omega_{max.strained} - \omega_{2D\ Te})/(\omega_{max.strained} - \omega_{bulk\ Te})$, where $\omega_{max.strained}$, $\omega_{bulk\ Te}$, and $\omega_{2D\ Te}$ are the peak values (in cm$^{-1}$) of $A^1$ phonon modes of maximum strained TeNF, bulk Te (relaxed) crystal, and that of the TeNF with a specific thickness, respectively. At a thickness of 12 nm, R is zero, however at a thickness of 60 nm, R grows to one as dislocations emerge, plastically relaxing the strain in their proximity.

## Text S3: Detailed XPS studies of polymorphic TeNFs:

To confirm whether a large PL signal purely originated from Te NFs, rather than from tellurium oxide species (i.e. TeO, $TeO_2$ etc), the chemical stoichiometry of as-prepared Te NFs was probed by x-ray photoemission spectroscopy (XPS). The comparative XPS spectra of bulk Te and as-prepared Te NFs are presented in Figure 6a. Deconvoluted Te 3d scan of Te NFs showing strong photoemissions at binding energies 571.2 eV and 582.9 eV are assigned to $3d_{5/2}$ and $3d_{3/2}$ spin-orbit doublet species of Te 3d, respectively, and suggest pure metallic phase of Te NFs.[2] No enlargement in oxidation peaks at binding energies 575.9 eV and 586.4 eV were observed as compared with bulk Te, which suggests a high degree of purity of as-prepared Te NFs. Figures 6b and c demonstrate O1s and C1s spectra of as-prepared Te NFs, respectively. Deconvolution of O1s core level spectrum after fitting the peaks by means of Lorentz–Gaussian (20–80%) function, while subtracting the Shirley background suggests doublet located at 531.5 eV and 530.1 eV. No evidence of strong oxidation was observed as the presence of tightly bound oxide and hydroxide is reported to exist typically at higher energies (≈532.6 eV).[3] XPS survey spectra of as-prepared Te NFs, bulk Te and bare silicon substrate are presented in Figure 6d. Owing to the small yield of Te NFs on Si substrate, the peak signal at characteristic binding energies of metallic Te is weak. A clear signal can be observed in the inset in the top right corner, highlighted by the red box.

## Text S4: COMSOL Multiphysics simulations for strain estimation

COMSOL Multiphysics software was used to simulate heating-induced stresses on ultrathin tellurium flakes. Within solid mechanics physics, a linear material model, together with its "Thermal Expansion" node, was used. The linear model is given by the following sets of equations:

$$\nabla \cdot \sigma = 0 \tag{1}$$

$$\sigma = C:\varepsilon \tag{2}$$

$$\varepsilon_{elastic} = \varepsilon - \varepsilon_{inelastic} = \varepsilon - \varepsilon_{ph} \tag{3}$$

$$\varepsilon = (1/2)[(\nabla u) + (\nabla u)T] \tag{4}$$

, where $\sigma$ is the stress tensor, C is the fourth order elasticity tensor, $\varepsilon_{ph}$ is the temperature-dependent strain and u is the displacement. External forces and initial stresses are zero.

The thermal strains at different cool-down temperatures were found by sweeping $0°C \leq T \leq 340°C$, using:

$$\varepsilon_{ph} = \alpha(T)(T - T_{ref}) \tag{5}$$

The reference temperature was $T_{ref} = 350$ °C, and $\alpha(T)$ is the coefficient of thermal expansion.

The coefficients of thermal expansion at elevated temperatures are taken as $16.8 \times 10^{-6}$ [1/K] and $5.15 \times 10^{-6}$ [1/K] for tellurium and sapphire (aluminum oxide), respectively[4,5]. Poisson's ratio is 0.33 and 0.305 (tellurium and sapphire, respectively). Young's modulus: $40 \times 10^9$ [Pa] and $356.5 \times 10^9$ [Pa] (tellurium and sapphire, respectively).

Quadratic serendipity shape functions were used for the simulations, and MUMPS was the solver of choice. The tellurium flakes were approximated to discs with a 40-nm radius (three different thicknesses: 5, 10 and 20 nm). Similarly, the underlying sapphire wafer was approximated to a cylinder with a height of 100 nm and a 50 nm radius. The mesh element sizes ranged between 5-10 nm.

*Text S5: Detailed Computational studies:*

All first-principles calculations are performed using the projector augmented wave method as implemented in Vienna ab-initio simulation package (VASP)[6] coupled with generalized gradient approximation (GGA) electron-electron interaction defined as Perdew-Burke-Ernzerhof (PBE) exchange-correlation functional[7]. The geometry of the system is optimized until Hellmann-Feynman force is converged to $10^{-4}$ eV/Å and the energy difference is converged to $10^{-6}$ eV[8,9]. A kinetic energy cutoff of 650 eV is used with a 25×20×1 and 13×20×1 k-points mesh for *t*-Se and *orth*-Se, and a vacuum space of 25 Å along the z-direction is taken to avoid interaction with other neighboring layers. First, we fully relaxed *t*-Se/*orth*-Se unit cell to obtain lattice parameters such a = 2.40/3.92Å, b = 4.63/4.95Å and c = 25 Å taken to avoid interaction with other neighboring layers, as shown in Fig. 4a (main text) structure of 3×3×1 *t*-Se/*orth*-Se supercell[10]. The lattice parameters with PBE are slightly larger while in good agreement with the available experimental results after computing with PBE exchange–correlation (XC) functional. The ELF is a position-dependent function having values from 0 to 1, where ELF = 1 corresponds to a completely localized electron[11]. Se bonds reveal the covalent nature of these layered structures.

To analyze the electronic properties of monolayer Te, the electron localization function (ELF) of mono-layer Te is presented in Fig. S9a, supporting information. The ELF is a position-dependent function having values from 0 to 1, where ELF = 1 corresponds to a completely localized electron. Te bonds reveal the covalent nature of these layered structures.[12]

Once the ground state electronic structures of a material are obtained, the optical properties can then be investigated, including the absorption spectra of dielectric functions and Joint density of states are investigated up to 5 eV photon energy to reveal the response of the material to high energy radiation. The optical absorption coefficient measures how much light at a particular wavelength (energy) penetrates the substance before it is absorbed. The monolayer Te highly anisotropic, which is one of the fundamental motives for checking these properties. A more extensive examination of the predicted absorption coefficients in the visible light band (1.65 - 3.26 eV) is required Fig. S9b, supporting information. In the low energy range, its absorption coefficient appears to be practically isotropic. For the absorption coefficient is somewhat anisotropic in ||x and ||y directions, as evidenced by the visually distinct curve forms in both directions.[13-15]

The frequency-dependent dielectric matrix was calculated within the independent-particle approximation.[16] Local field effects and many-body effects were not considered, which was proven to be adequate to quantify the optical contrast for mono-layer Te.[17] The frequency-dependent complex dielectric function $\varepsilon(\omega)$ where, $\varepsilon(\omega) = \varepsilon_1(\omega) + i\varepsilon_2(\omega)$, can be used to determine the linear optical characteristics of semiconductors, where and are the $\varepsilon_1(\omega)$ real and $\varepsilon_2(\omega)$ imaginary parts of the dielectric function, and $(\omega)$ is the photon frequency. Where, absorption coefficient $\alpha(\omega)$, can be calculated from the real $\varepsilon_1(\omega)$ and the imaginary $\varepsilon_2(\omega)$ parts:

$$\alpha(\omega) = \frac{\sqrt{2}\omega}{c}\left[\sqrt{\varepsilon_1^2 + \varepsilon_2^2} - \varepsilon_1\right]^{\frac{1}{2}} \quad (2)$$

To disentangle the role of the matrix elements, we computed the Joint Density of States (JDOS) $J(\omega)$ defined by where the sum is over the valence and conduction states at the supercell Γ-point. A strong optical contrast is present between the unstrained and strained and the mostly 1-D chain-like at high strain up to -4.6%, although the bonding geometry is quite similar. In Te layer, both the change in the optical matrix elements and the change in the $(J(\omega)/\omega^2)$ contribute to the optical contrast between the 1-D chain or flake. where the JDOS is the joint density of states defined as

$$J(\omega) = 2 \sum_{v,c,k} w_k \, \delta(E_c - E_v - \omega), \qquad (3)$$

Where $E_v$ and $E_c$ indicate the energies of the states in the valence and conduction bands, respectively. Between 1 and 2 eV in the energy range of the prominent peak of the absorption spectrum is greater, and the JDOS in the strained monolayer is much stronger than in the unstrained phase. The optical absorption spectrum can be considered proportional to this JDOS, corresponding to the density of possible independent transitions (Figure 4(b)).

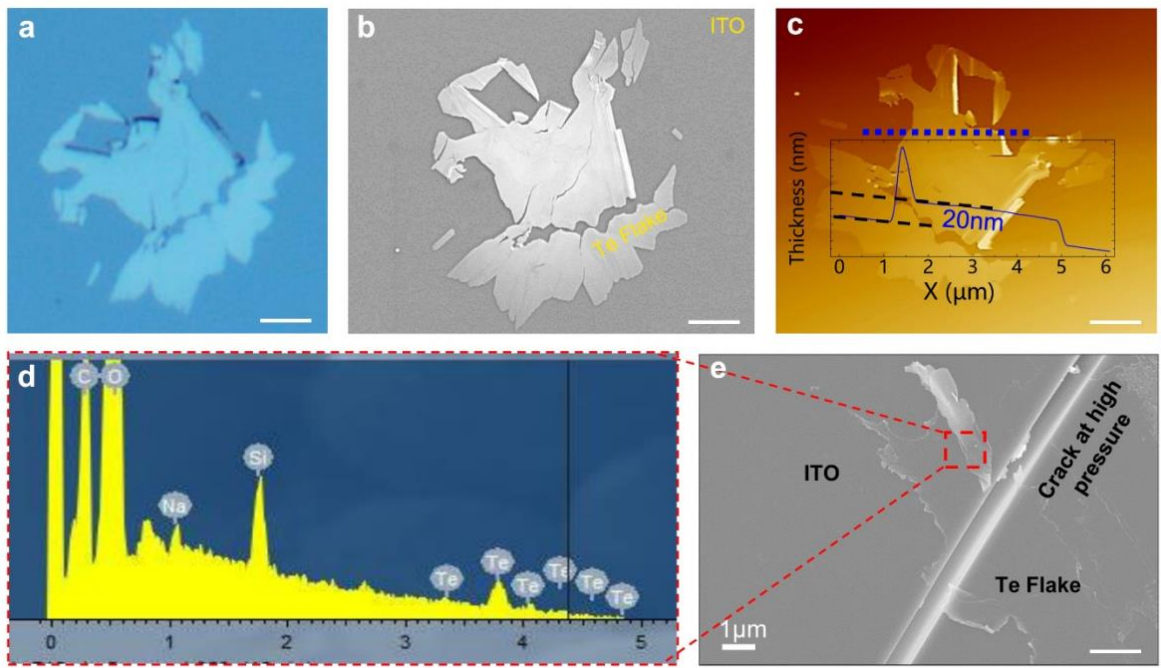

**Fig. S1. a,** The optical microscopy image of a TeNF fabricated on ITO substrate. **b,** Field emission scanning electron microscopy (FESEM) image of the same TeNF/ITO. **c,** Atomic force microscopy image of the identical TeNF/ITO with its height profile. The thickness was estimated to be 20nm along the blue dotted line trace. **d,** EDX map of TeNF acquired from Fig. S1e, showing the strong tellurium signal. **e,** The SEM image featuring the crack propagation in ITO through a Te flake because of high pressure. The scale bar in all the figures is 5μm.

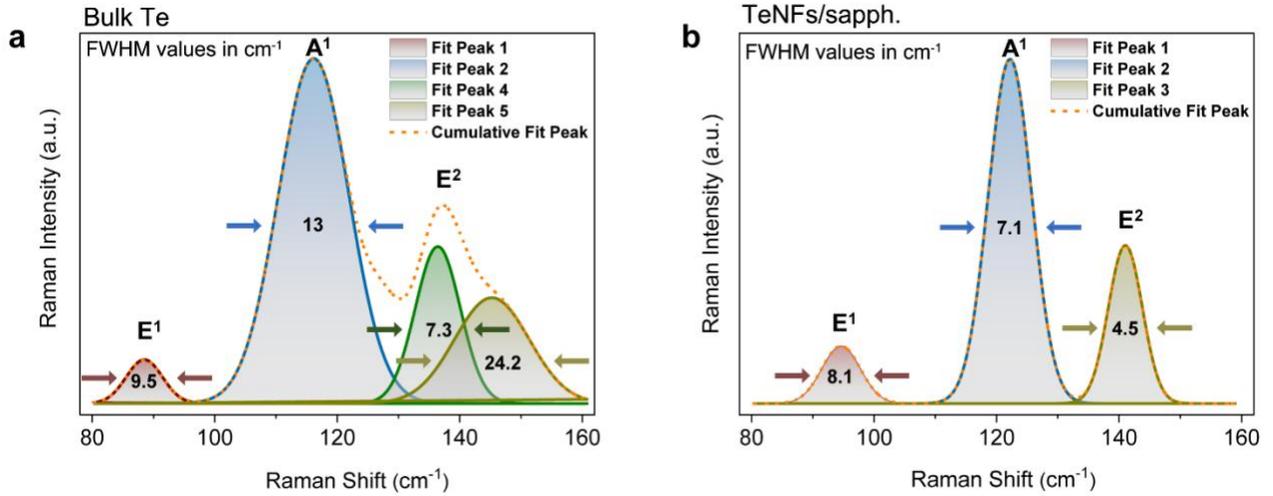

**Fig. S2. a,** A deconvoluted μ-Raman spectrum obtained from bulk Te. **b,** A deconvoluted μ-Raman spectrum obtained from TeNFs/sapph. Peak fitting was performed using Gaussian fitting to estimate the FWHM (Γ), which is numerically presented with in the corresponding peaks.

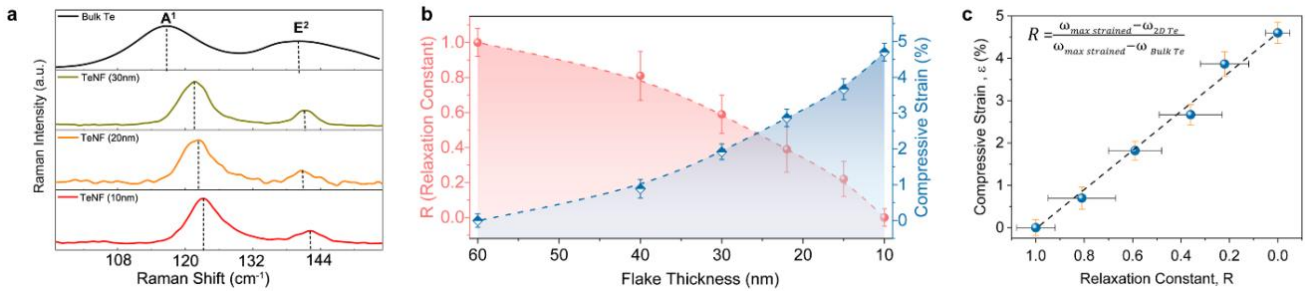

**Fig. S3. a,** Thickness-dependent μ-Raman spectra displaying a consistent blueshift in $A^1$ and $E^2$ vibrational modes with decreasing flake thicknesses. **b,** Correlation between relaxation constant and compressive strain, as a function of flake thicknesses. **c,** Direct relation between the compressive strain and relaxation constant (R), where the inset is the formula used to compute the R from the μ-Raman data.

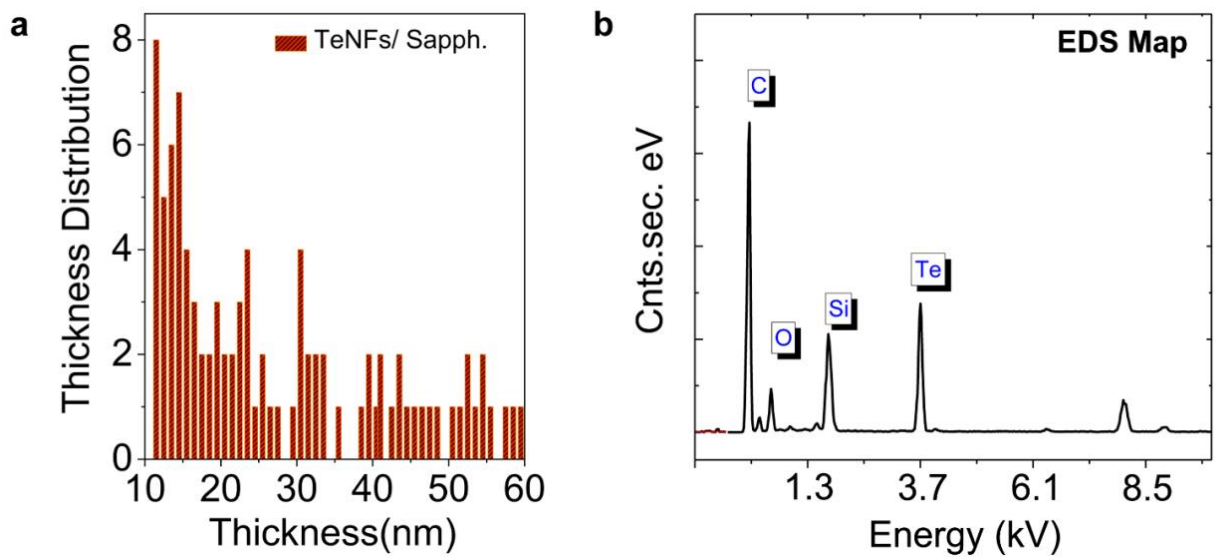

**Fig. S4**. **a,** Thickness distribution histogram acquired by the AFM data of 100 thinnest TeNFs. **b,** EDS map of TeNFs acquired using TEM confirms strong signal for tellurium.

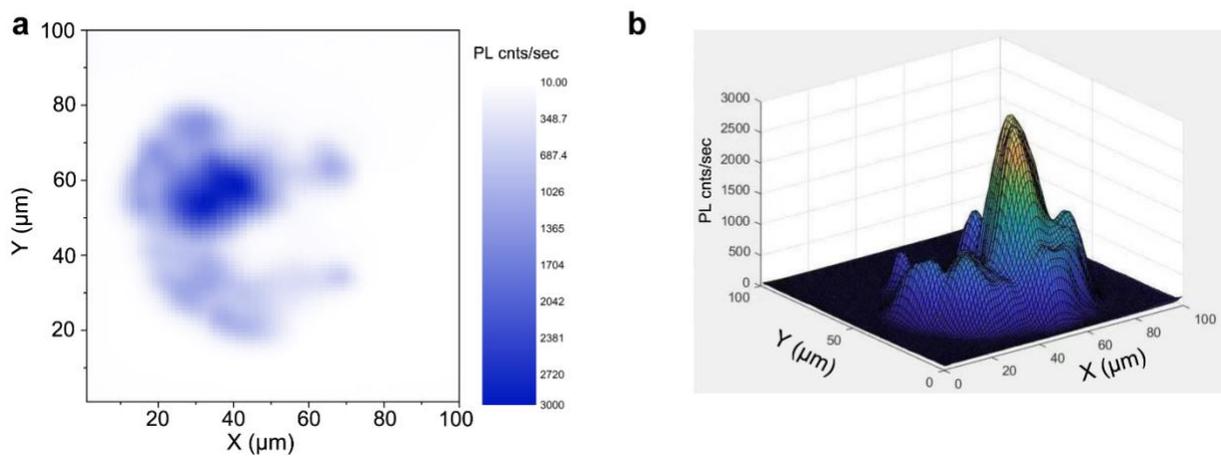

**Fig. S5**. **a,** A 10μm×10 μm PL map of a TeNF/sapphire showing the spatially distributed photoemission. **b,** A 3D plot of the PL map demonstrating strong PL localization at the center of the flake.

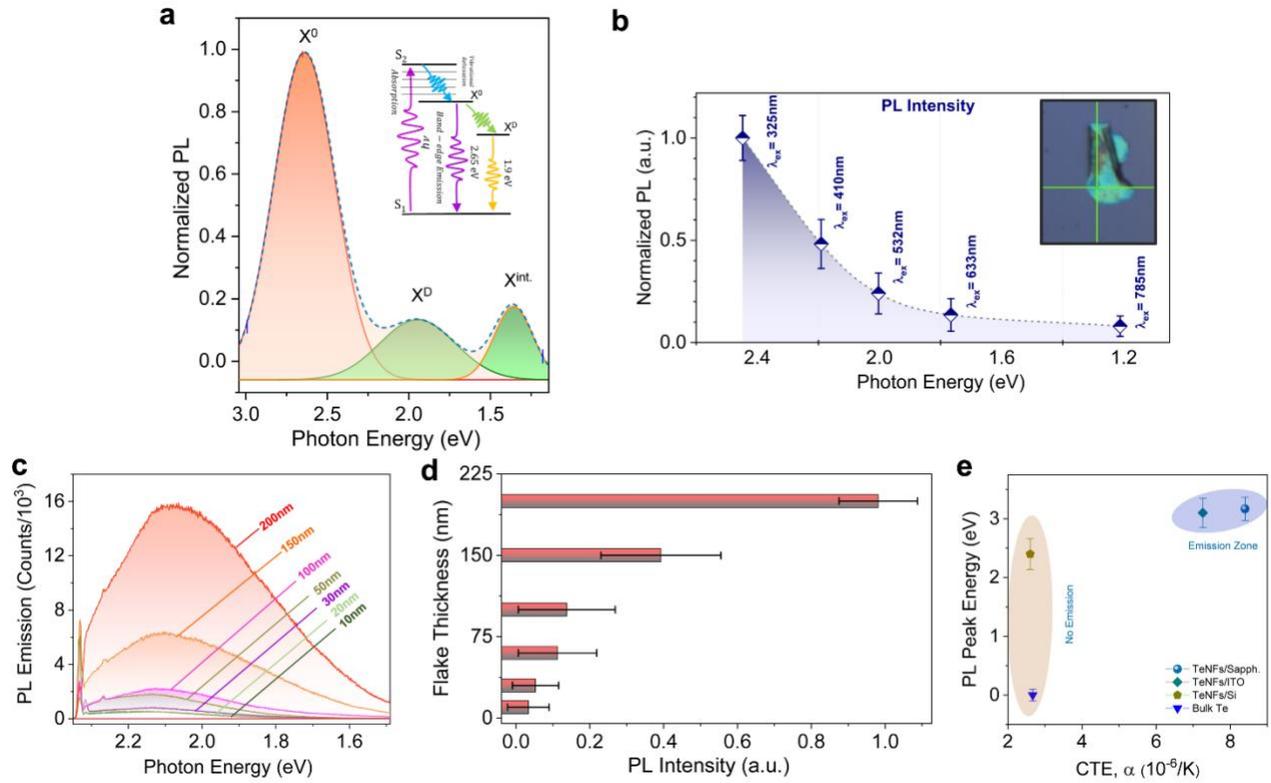

**Fig. S6. a**, Deconvoluted PL spectrum of TeNF/sapphire showing the peak consisting of three singlets ($X^0$, $X^D$, and $X^{Int.}$). The inset shows the schematic of inner-shall transitions during the radiative excitation recombination process. **b,** The variation in PL intensity with various excitation wavelengths. **c,** Thickness-dependent PL response of TeNFs with using 532 nm laser. **d,** A bar graph showing the quantitative variation in normalized PL intensity with flake thicknesses. **e,** The PL peak energies as a functions of substrate's CTE.

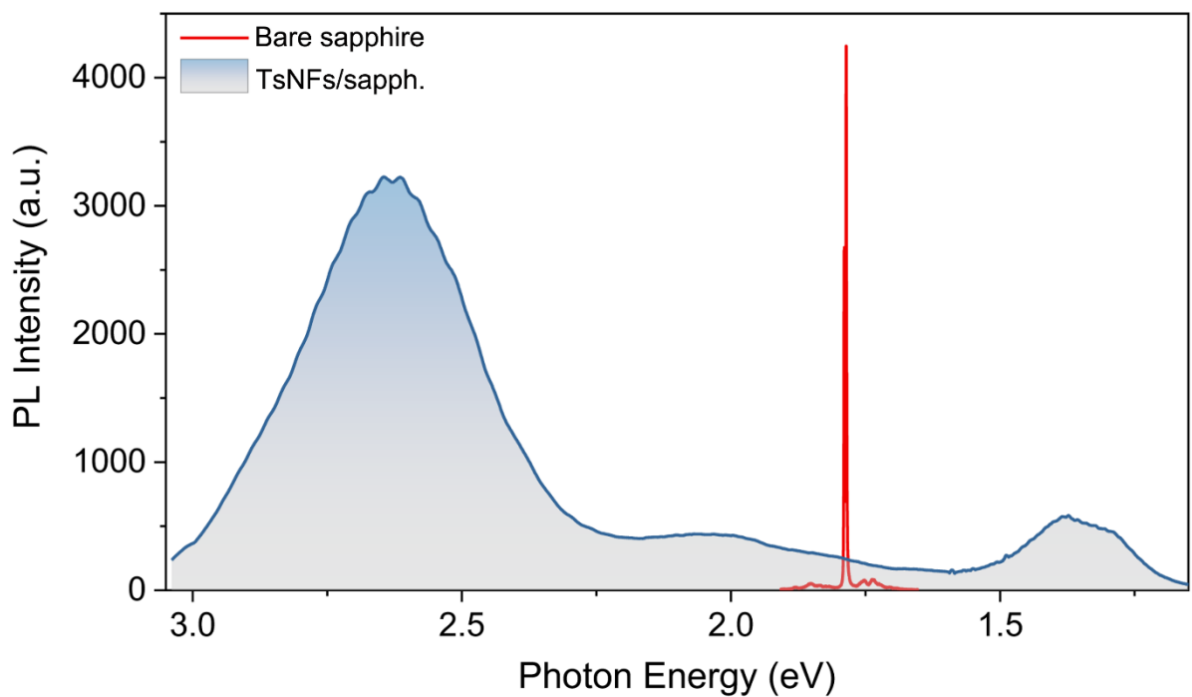

**Fig. S7**. The comparative PL spectra of bare sapphire (in red) and TeNFs/sapphire (in blue), confirming no background PL signal from sapphire at around 1.79 eV.

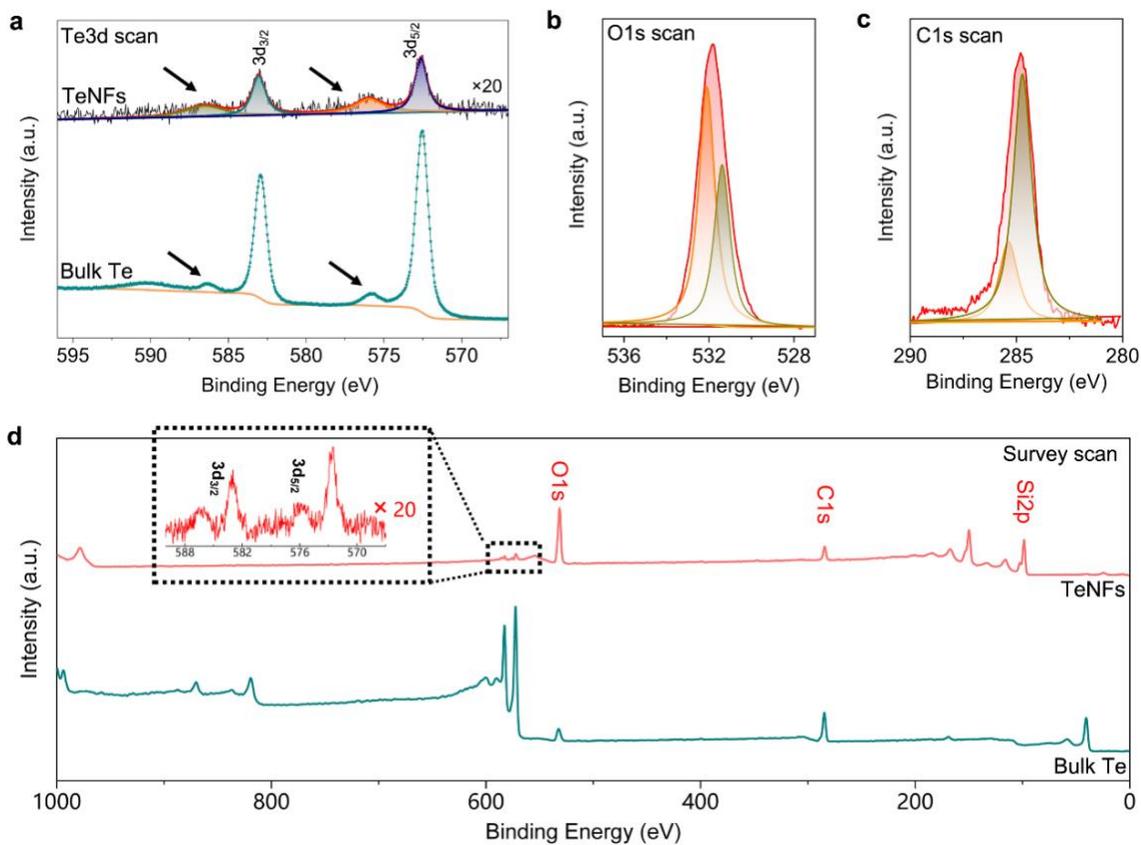

**Fig. S8. a,** Te3d scan of bulk Te (powder) and Te NFs fabricated on Si substrate to avoid elemental complexities offered by sapphire and ITO substrates. **b,** O1s scan of as-prepared TeNFs **c,** C1s spectra of as-prepared TeNFs. **d,** Survey scans of bulk Te and TeNFs. The inset shows the 20 times magnified spectra of Te NFs acquired from the area highlighted by the black rectangle.

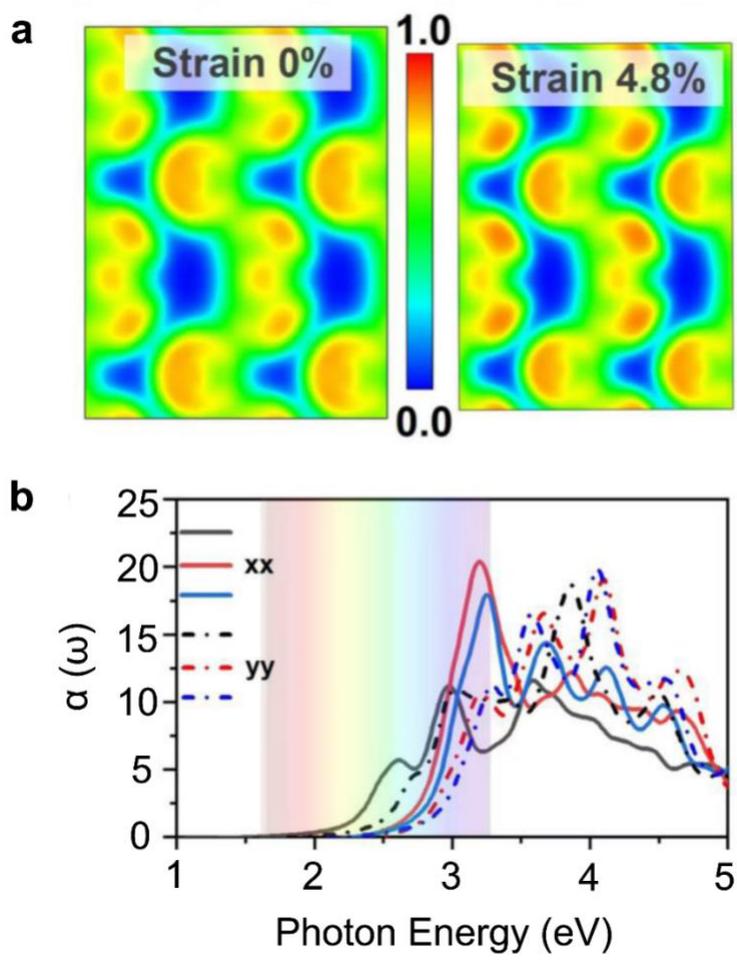

**Fig. S9. a**, Electron localization function (ELF) of unstrained and strained Te flakes. **b,** Absorption spectra of TeNFs, obtained by DFT calculations.

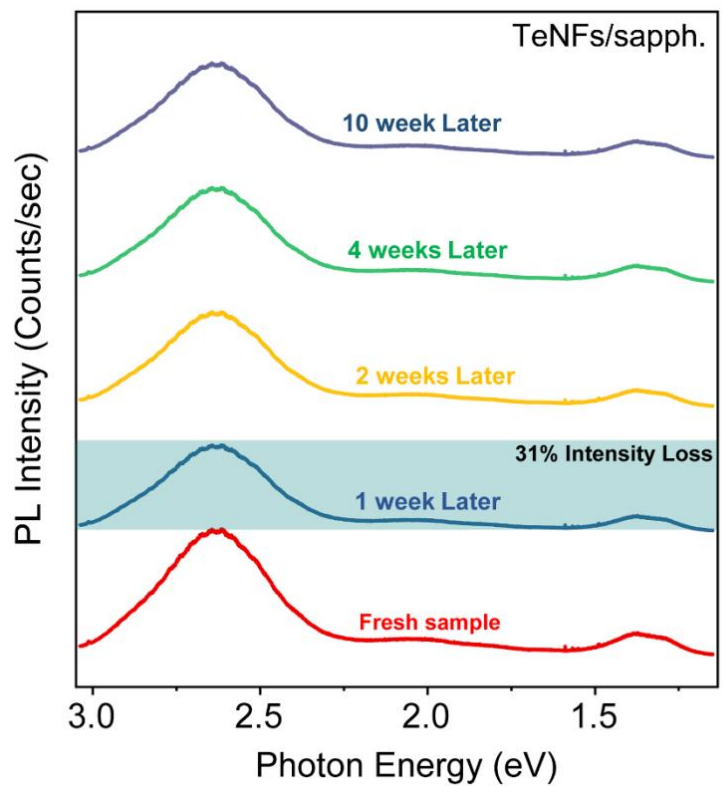

**Fig. S9.** Time-dependent of PL spectra of TeNFs/sapphire showing consistent PL response after 10 weeks of fabrications. This also confirms long-term bandgap retention in TeNFs.

| Material | Straining Method | Max. Strain (%), strain type | Bandgap Modulation (meV) | $E_g$ Modulation Rate (meV/%) | Ref. |
|---|---|---|---|---|---|
| $WSe_2$ | Mechanical Bending | 1.5, Volatile | 110 | 73.3 | Strain-induced indirect to direct |
| $WSe_2$ | Substrate Induction | 0.2-1.0, Non-volatile | 140 | 140 | Strain-engineered growth of two-dim |
| $MoS_2$ | Mechanical Bending | 0.2, volatile | 60 | 300 | Exceptional tunability of band energy |
| $MoS_2$ | Mechanical Bending | 4, volatile | 360 | 90 | Bandgap tuning of two-dimensional materials by sphere diameter |
| $WSe_2$ | Mechanical Bending | 1.4, volatile | 76 | 54 | Reversible uniaxial strain tuning in atomically thin WSe2 |
| $MoS_2$ | Mechanical Bending | 2.2, volatile | 260 | 120 | Bandgap engineering of strained monolayer |
| $MoS_2$ | Mechanical Bending | 0.8, volatile | 68.8 | 86 | Strain tuning of optical emission |
| graphene | Water adsorption | -----, volatile | 2.6 | --- | Tunable bandgap in graphene by the controlled adsorption |
| $MoS_{2(1-x)}Se_{2x}$ | Chemical doping | -----, volatile | 130 | --- | Growth of large-area 2D MoS2 (1-x) Se2x semiconductor |
| $MoSe_2$ | Chemical doping | -----, volatile | 102 | -- | Growth of alloy MoS2 x Se2 (1–x) nanosheets |
| graphene | High-pressure | -----, volatile | 2.5 | --- | Large bandgap of pressurized trilayer |
| $MoS_2$ | Mechanical Bending | 0.52, volatile | 35 | 70 | Experimental demonstration of continuous electronic structure |
| Halide Perovskite | Epitaxial growth | -2.4, Non-volatile | 35 | 14.5 | Strain engineering and epitaxial stabilization of halide perovskites |
| Tellurium | Thermomechanical compression | -4.6, Non-volatile | 2650 | 580 | This Work |